\documentclass[prl,aps,twocolumn,superscriptaddress]{revtex4-1}
\pdfoutput=1

\usepackage{amssymb}
\usepackage{epsfig}
\usepackage{amsmath}
\usepackage{subfigure}
\usepackage{graphicx}
\usepackage{textcomp}
\usepackage{url}
\usepackage{float}
\usepackage{color}
\usepackage{cases}
\usepackage[normalem]{ulem}

\usepackage[titletoc,title]{appendix}

\usepackage[colorlinks=true, urlcolor=blue, anchorcolor=blue, citecolor=blue,filecolor=blue,linkcolor=blue,menucolor=blue%,pagecolor=blue
]{hyperref}

\usepackage{color}

\begin{document}
\title{Negative autocorrelations of disorder strongly suppress thermally activated particle motion in short-correlated quenched Gaussian disorder potentials}

\author{Baruch Meerson}
\email{meerson@mail.huji.ac.il}
\affiliation{Racah Institute of Physics, Hebrew University of
Jerusalem, Jerusalem 91904, Israel}

\begin{abstract}
We evaluate the mean escape time of overdamped particles over potential barriers in short-correlated quenched Gaussian disorder potentials in one dimension at low temperature. The thermally activated escape  is very sensitive to the form of the \emph{tail} of the potential barrier probability distribution. We evaluate this tail by using the optimal fluctuation method. For monotone decreasing autocovariances we reproduce the tail obtained by Lopatin and Vinokur (2001).  However, for nonmonotonic autocovariances of the disorder potential we show that the tail changes. It is much higher when the disorder potential exhibits negative autocorrelations, and it is much lower when the  autocovariance is nonmonotonic but everywhere positive. This leads to an \emph{exponential} increase or decrease, respectively, of the mean escape time.

\end{abstract}

\maketitle
\nopagebreak

\textit{Introduction.} Overdamped particle motion in short-correlated quenched disorder potential  in the presence of thermal noise is an important paradigm in a multitude of applications. These range from diffusive transport of electrons,
holes and excitons in disordered metals or semiconductors \cite{solidstate1,solidstate2}, to supercooled liquids and glassy matrices \cite{Bassler,glass1,glass2,glass3} and
to DNA macromolecules in living systems \cite{bio1,bio2,bio3}. Colloidal systems in quenched random potentials, created by laser light, have recently become
experimentally available \cite{G1,G2,G3,G4,speckle}. Starting from the groundbreaking works  of de Gennes and Zwanzig \cite{DeGennes,Zwanzig}, different aspects of this basic paradigm have attracted much interest from theorists \cite{LV,anomalous1,anomalous2,G3,anomalous4,Wilkinson}.

In the simplest one-dimensional setting,  overdamped particle motion in short-correlated quenched disorder potential $V(x)$ in the presence of thermal noise can be described by the Langevin equation
\begin{equation}\label{Langevin}
\dot{x} = -\frac{d V(x)}{dx}+ \sqrt{2D} \xi(t)\,.
\end{equation}
Here $D$ is the particle diffusion coefficient in the absence of disorder potential, and $\xi(t)$ is a delta-correlated Gaussian noise with zero mean and $\langle\xi(t) \xi(t^{\prime}) \rangle = \delta(t-t^{\prime})$. We suppose in the following that the potential $V(x)$ is statistically stationary and normally distributed with zero mean and variance $C(0)$. At low temperature, $D \to 0$, a particle rapidly settles down in a local potential minimum, but ultimately escapes by overcoming a potential barrier $\Delta V$, which is defined as the minus difference between the heights of the minimum and one of the two adjacent maxima of $V(x)$. The average escape time over one such potential barrier (where the averaging is performed over the thermal noise) is well known \cite{Kramers},
\begin{equation}\label{Kramerstime}
\bar{T} \sim \exp \left(\frac{\Delta V}{D}\right)\,,
\end{equation}
up to a pre-exponential factor that we will not be interested in. Our main objective is to determine the mean escape time $\langle\bar{T}\rangle$, where the additional averaging is performed over realizations of the disorder potential. At $D\to 0$, $\langle\bar{T}\rangle$ is strongly affected by the \emph{tail} of the probability distribution of the barrier heights which we will call $\mathcal{P}(\Delta V)$. This distribution tail is expected to behave as
\begin{equation}\label{tailgeneral}
\mathcal{P}(\Delta V \to \infty) \sim \exp\left[-S(\Delta V)\right]\,,
\end{equation}
where $S(\Delta V)$ is an \textit{a priori} unknown large-deviation function. Once $S(\Delta V)$ is determined, one can average
the mean escape time~(\ref{Kramerstime}) over the distribution tail (\ref{tailgeneral}):
\begin{equation}\label{Tdoubleav1}
\langle \bar{T} \rangle \sim \int d \Delta V \exp\left[\frac{\Delta V}{D}-S(\Delta V)\right]\,.
\end{equation}
The integral can be evaluated by the Laplace's method:
\begin{equation}\label{Tdoubleav2}
\langle \bar{T} \rangle \sim \exp\left[\frac{\Delta V_*}{D}-S(\Delta V_*)\right]\,,
\end{equation}
where the saddle point $\Delta V_*$ is determined from the equation
\begin{equation}\label{saddle}
D \frac{d S(\Delta V_*)}{d \Delta V_*} = 1\,.
\end{equation}
Now we proceed to the evaluation of the large-deviation function $S(\Delta V)$.

\textit{Optimal fluctuation method.} To complete the statistical description of our short-correlated Gaussian  disorder potential $V(x)$, we introduce the autocovariance
\begin{equation}\label{correlatorgeneral}
 C(x-x^{\prime})=\langle V(x) V(x^{\prime})\rangle
\end{equation}
and rewrite it as $C(x-x^{\prime})=A \,\kappa(x-x^{\prime})$, where
$\kappa(z)$,  an even function of $z$, is normalized to unity,
\begin{equation}\label{normalization}
\int_{-\infty}^{\infty} \kappa(z) dz =1\,,
\end{equation}
and  $A \kappa(0)=C(0)>0$ is the variance. (For brevity, we will use the word ``covariance" instead of the ``autocovariance" in the following.) We will also need the kernel $K(x-x^{\prime})$, inverse to $\kappa(x-x^{\prime})$. It is defined by the relation
\begin{equation}\label{inverse}
\int_{-\infty}^{\infty} dx^{\prime\prime} \,K(x-x^{\prime\prime}) \,\kappa(x^{\prime}-x^{\prime\prime}) =\delta(x-x^{\prime}).
\end{equation}
By virtue of Eq.~(\ref{inverse}), $K(z)$ is also normalized to unity: $\int_{-\infty}^{\infty} K(z) dz =1$. The knowledge of the inverse kernel $K$ enables one to write down the statistical weight of a given realization of a normally distributed stationary short-correlated random field $V(x)$ \cite{Zinn-Justin}.  Up to normalization,  the statistical weight is equal to  $\exp\{-\mathcal{S}[V(x)]\}$, where the action functional $\mathcal{S}[V(x)]$ is
\begin{equation}\label{actiongeneral}
\mathcal{S}[V(x)] = \frac{1}{2A} \int_{-\infty}^{\infty} dx \int_{-\infty}^{\infty} dx^{\prime} K(x-x^\prime) V(x) V(x^\prime).
\end{equation}
The distribution tail of $\mathcal{P}(\Delta V)$ corresponds to atypically large $\Delta V$ which are dominated by the \emph{optimal}, that is most likely, configuration of the potential $V(x)$ conditioned on this $\Delta V$.  This is the essence of the optimal fluctuation method (OFM) (also called the instanton method) that we employ here, following an early work of Lopatin and Vinokur \cite{LV}.   The OFM boils down to a minimization of the action functional (\ref{actiongeneral}) with respect to $V(x)$ subject to constraints that we now specify. Without loss of generality, we can place the minimum and maximum of $V(x)$ at $x=-L$ and $x=L$, respectively. Overall, assuming that the optimal solution $V(z)$ is smooth, we demand
\begin{eqnarray}
  &&V(x=L)-V(x=-L)= \Delta V\,,
  \label{constraints}\\
 &&\frac{dV}{dx}\left(x=-L\right)=0\,, \quad \frac{d^2V}{dx^2}\left(x=-L\right)>0\,,\label{defminusL}\\
 &&\frac{dV}{dx}\left(x=L\right)=0\,,\quad \frac{d^2V}{dx^2}\left(x=L\right)<0\,.\label{defL}
\end{eqnarray}
In addition, there must be no other extrema inside the interval $|x|<L$, and the minimization is over all possible values of $L$.

It is convenient to rewrite Eq.~(\ref{constraints}) as an integral constraint:
\begin{equation}\label{HHH}
\int_{-\infty}^{\infty} V(x) \left[\delta(x-L)- \delta(x+L)\right]dx =\Delta V.
\end{equation}
Now we can minimize the functional
\begin{eqnarray}
 s_{\lambda}[V(x)]&=& \frac{1}{2}\int_{-\infty}^{\infty} dx \left\{\int_{-\infty}^{\infty} dx^\prime  K(x-x^\prime) V(x) V(x^\prime) \right. \nonumber\\
  &-& \left. \lambda V(x) \left[\delta(x-L)- \delta(x+L)\right]\right\}\,,
\label{functional}
\end{eqnarray}
where $\lambda$ is a Lagrange multiplier to be ultimately expressed through $\Delta V$ from Eq.~(\ref{constraints}). The variation $\delta s_{\lambda}$ must vanish,
\begin{eqnarray}
  \delta s_{\lambda} &=& \int_{-\infty}^{\infty} dx \,\delta V(x) \left\{\int_{-\infty}^{\infty} dx^{\prime} K(x-x^\prime) V(x^\prime)\right. \nonumber \\
  &-& \left.\frac{\lambda}{2} \left[\delta(x-L)- \delta(x+L)\right] \right\} =0,
  \label{B90}
\end{eqnarray}
leading to the linear equation
\begin{equation}\label{B100}
\!\int_{-\infty}^{\infty} dx^\prime\,K(x-x^\prime)V(x^\prime) \!= \!\frac{\lambda}{2} \left[\delta(x-L)- \delta(x+L)\right],
\end{equation}
which, for given $L$ and $\lambda$, has a unique solution. Comparing Eqs.~(\ref{inverse}) and~(\ref{B100}), we can easily guess this solution:
\begin{equation}\label{V1}
V(x) = \frac{\lambda}{2}\left[\kappa(x-L)-\kappa(x+L)\right]\,,
\end{equation}
so what remains is to determine $L$ and express $\lambda$ through the rest of parameters.
$V(x)$ is an odd function of $x$, and we demand that it have its maximum  at $x=L$. The condition $dV/dx(x=L)=0$ yields
\begin{equation}\label{maximumcondition}
\frac{d\kappa(x)}{dx}\Big|_{x=0} - \frac{d\kappa(x)}{dx}\Big|_{x=2L}=0.
\end{equation}
For smooth covariances the first term in Eq.~(\ref{maximumcondition}) vanishes. The behavior of the second term depends on whether the covariance function $\kappa(x)$ is a monotone decreasing function of $x>0$ or not, so we will consider these two cases separately.

\begin{figure} [ht]
\includegraphics[width=0.30\textwidth,clip=]{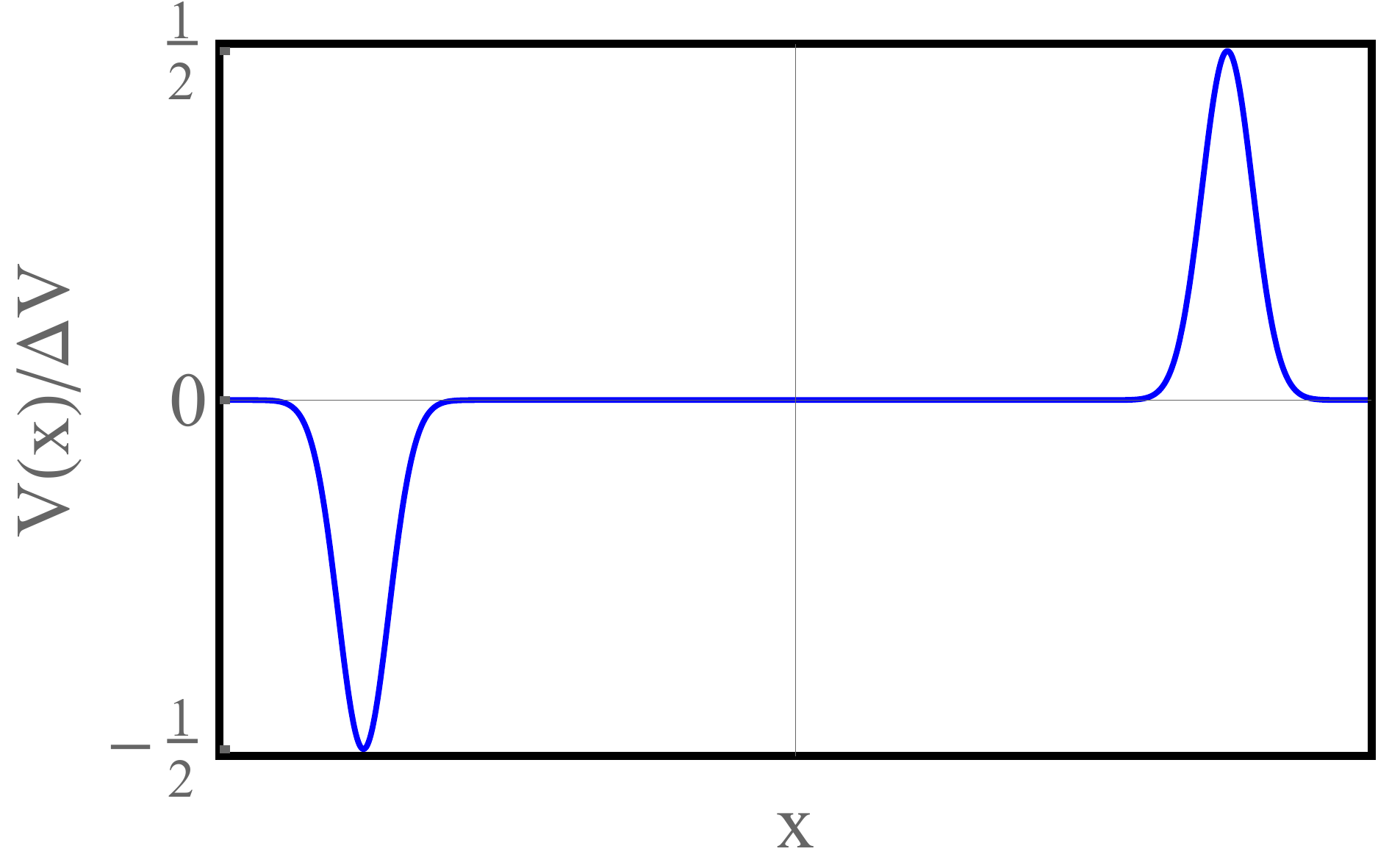}
\caption{Optimal configuration of the disorder potential $V(x)$, conditioned on a large value of the potential barrier $\Delta V$ for a monotone decreasing covariance, see Eq.~(\ref{V2}).}
\label{twospikes}
\end{figure}

\textit{Monotone-decreasing covariance.} In this case the second term in Eq.~(\ref{maximumcondition}) vanishes only in the limit of $L\to \infty$. Therefore, according to Eq.~(\ref{V1}), the most probable configuration of the potential $V(x)$, conditioned on a large $\Delta V$,
\begin{equation}\label{V2}
V(x) = \frac{\Delta V}{2 \kappa(0)}\left[\kappa(x- L)-\kappa(x+L)\right] , \quad  L\to \infty,
\end{equation}
consists  of two essentially \emph{independent} ``spikes": a positive spike with the height $(\lambda/2) \kappa(0)=\Delta V/2$ at $x=L$ and a negative spike with height $-(\lambda/2) \kappa(0)= -\Delta V/2$ at $x=-L$ \cite{otherconditions}.
An example of this configuration is shown Fig. \ref{twospikes}.  Remarkably, the spikes of optimal $V(x)$ have the same functional form as the covariance function itself.

Using Eq.~(\ref{V2}), we can evaluate the action (\ref{actiongeneral}). The spike and anti-spike give equal contributions and, using Eq.~(\ref{inverse}), we obtain the large deviation function
\begin{equation}\label{actionresultgen}
S(\Delta V) = \frac{\Delta V^2}{4A \kappa(0)} = \frac{\Delta V^2}{4 C(0)} \,.
\end{equation}
This result  is independent of the form of the (monotone-decreasing) covariance  of the disorder potential: it depends only on the variance $C(0)$.
The corresponding tail of the barrier height distribution,
\begin{equation}\label{probmonotone}
 \mathcal{P}(\Delta V) \sim \exp\left[-\frac{\Delta V^2}{4C(0)}\right]\,,\quad \Delta V \to \infty\,,
\end{equation}
perfectly agrees with that of Lopatin and Vinokur \cite{LV,aboutLV}. Now, using Eqs.~(\ref{saddle}) and (\ref{actionresultgen}), we obtain the saddle point $\Delta V_* = 2 C(0)/D$ which, by virtue of Eq.~(\ref{Tdoubleav2}), leads to
\begin{equation}\label{Taveragemon}
\langle \bar{T} \rangle \sim \exp \left[\frac{C(0)}{D^2}\right]\,,\quad D\to 0\,.
\end{equation}
Because of the $1/D^2$ factor in the denominator inside the exponent, the mean escape time is extremely long \cite{DeGennes,Zwanzig}. Importantly,  for monotone decreasing covariance $C(z)$, the mean escape time is determined by the
variance of the disorder potential and is insensitive to the form of the covariance.

\begin{figure} [ht]
\includegraphics[width=0.27\textwidth,clip=]{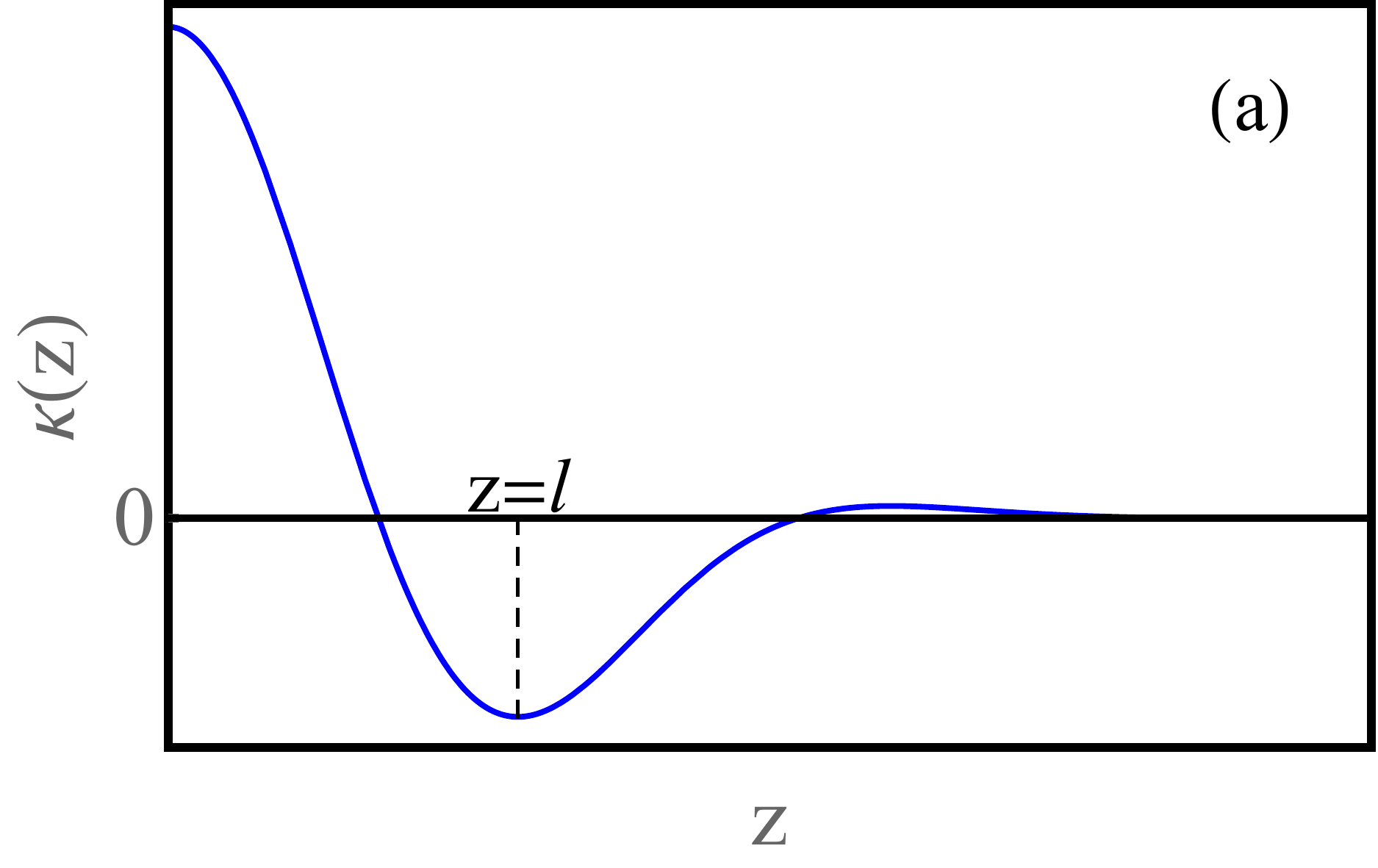}
\includegraphics[width=0.27\textwidth,clip=]{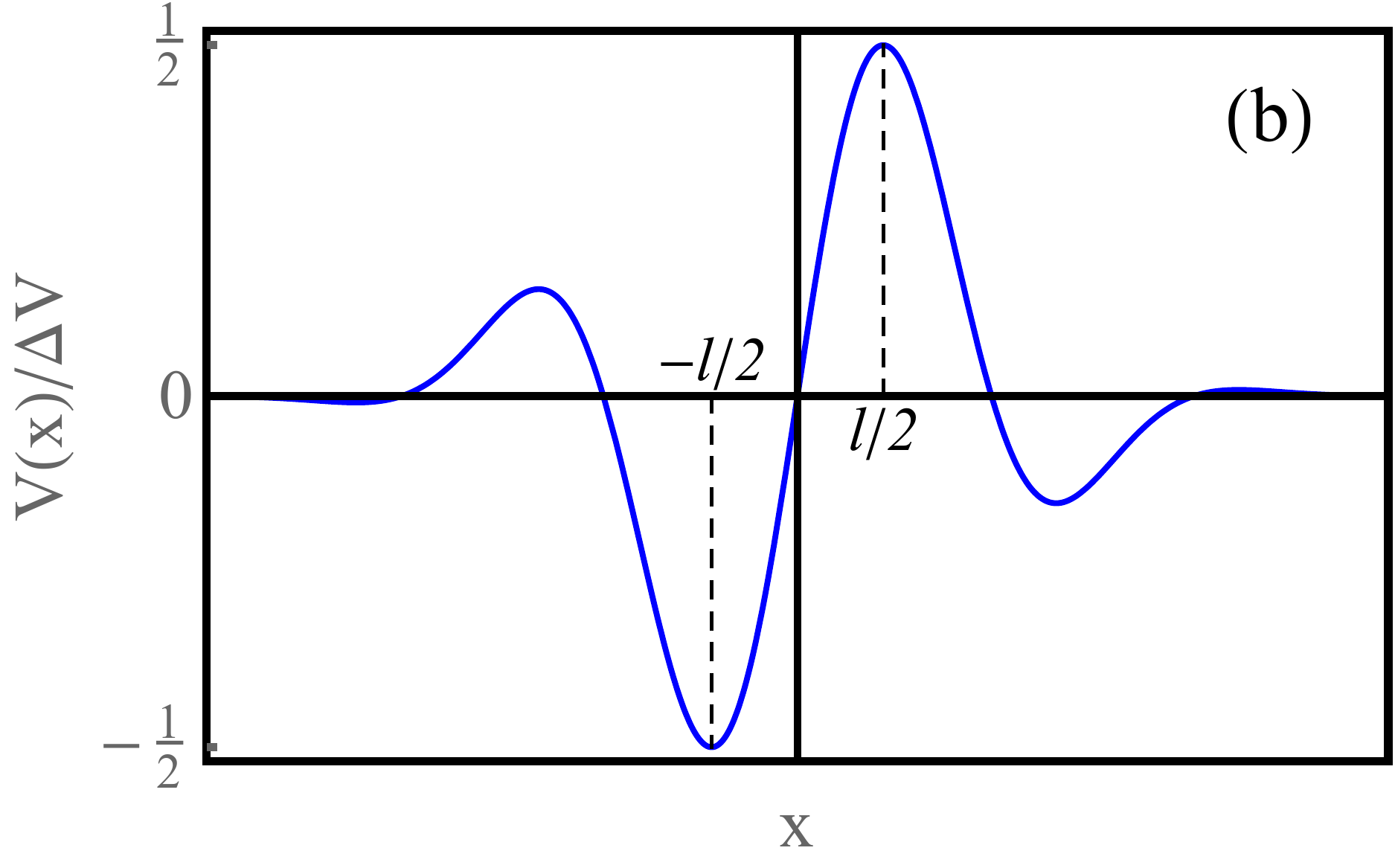}
\caption{A nonmonotonic covariance, exhibiting a region of negative correlations (a), and
the optimal configuration of the disorder potential $V(x)$ in this case (b).}
\label{covnegative}
\end{figure}

\begin{figure} [ht]
\includegraphics[width=0.27\textwidth,clip=]{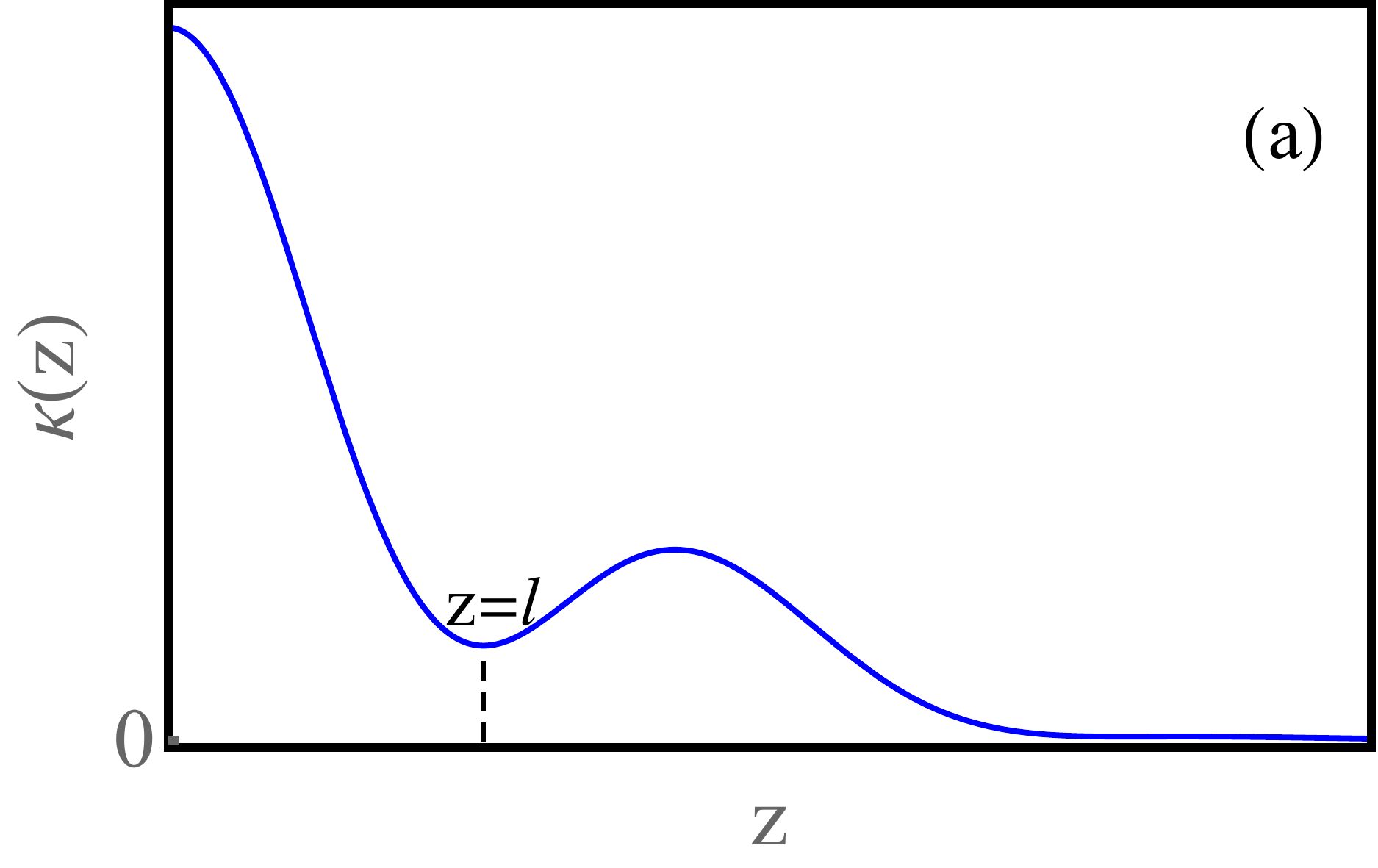}
\includegraphics[width=0.27\textwidth,clip=]{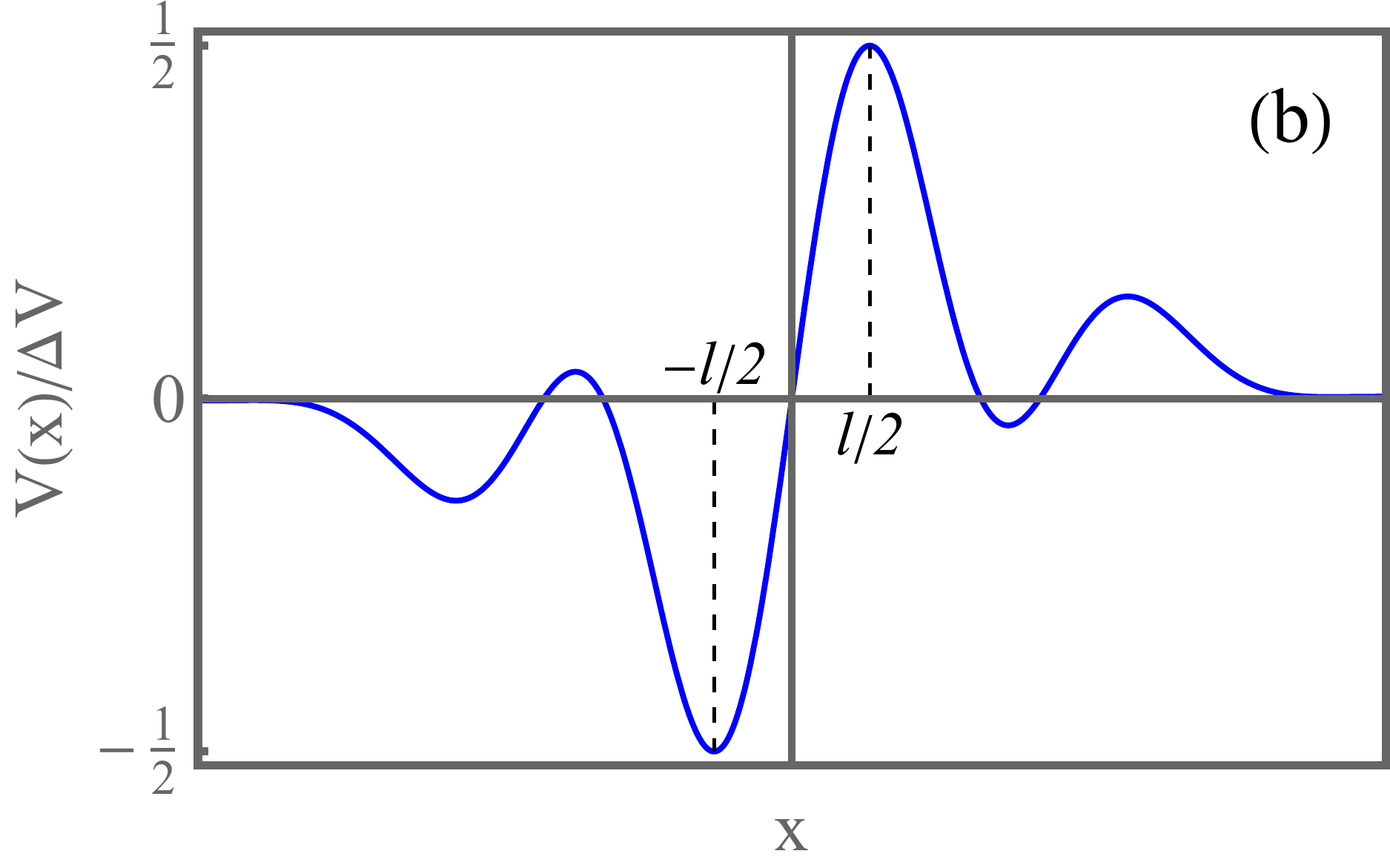}
\caption{A nonmonotonic but positive covariance (a) and
the optimal configuration of the disorder potential $V(x)$ in this case (b).}
\label{covpositive}
\end{figure}

\textit{Nonmonotonic covariances.} Now we apply Eq.~(\ref{maximumcondition}) to a nonmonotonic covariance \cite{limitation}. We continue to assume that the disorder potential $V(x)$ is smooth, so that $d\kappa/dx(x=0)=0$. A nonmonotonic $\kappa(x)$ implies existence of at least one minimum at $x>0$. Let us denote by $x=\ell>0$ the minimum point of $\kappa(x)$ closest to $x=0$.  Now, to satisfy Eq.~(\ref{maximumcondition}) and the rest of conditions we must set $L=\ell/2$. This is in contrast to the case of monotonic covariance, where we could only set $L\to \infty$. We determine  $\lambda$ from
Eq.~(\ref{constraints}) and arrive at the following optimal configuration of $V(x)$:
\begin{equation}\label{Vnonmonotonic}
V(x) = \frac{\Delta V}{2}\,\frac{\kappa\left(x-\frac{\ell}{2}\right)-\kappa\left(x+\frac{\ell}{2}\right)}
{\kappa(0)-\kappa\left(\ell\right)} .
\end{equation}
Plugging Eq.~(\ref{Vnonmonotonic}) into Eq.~(\ref{actiongeneral}), splitting the integral into four integrals, and making use of Eq.~(\ref{inverse}) and of the symmetry of the covariance $\kappa(-z)=\kappa(z)$, we obtain after some algebra
\begin{equation}\label{Snonmonotonic}
S(\Delta V) = \frac{\Delta V^2}{4A \left[\kappa(0)-\kappa(\ell)\right]}\,.
\end{equation}
This result looks more interesting than Eq.~(\ref{probmonotone}), where only the variance $A\kappa(0)=C(0)$ contributes.
Using Eq.~(\ref{Snonmonotonic}), we obtain the tail of $\mathcal{P}(\Delta V)$:
\begin{equation}\label{probnonmonotonic}
 \mathcal{P}(\Delta V) \sim \exp\left\{-\frac{\Delta V^2}{4A \left[\kappa(0)-\kappa(\ell)\right]}\right\}\,,\quad \Delta V \to \infty\,.
\end{equation}
The saddle point $\Delta V_*$, described by Eqs.~(\ref{saddle}) and (\ref{Snonmonotonic}), is now
$\Delta V_* = 2 A [ \kappa(0)-\kappa(\ell)]/D$. Then, using Eq.~(\ref{Tdoubleav2}), we arrive at
\begin{eqnarray}
\langle \bar{T}\rangle&\sim & \exp \left\{\frac{A [\kappa(0)-\kappa(\ell)]}{D^2}\right\}  \nonumber \\
&=& \exp \left[\frac{C(0)-C(\ell)}{D^2}\right]\,,\quad D\to 0\,.
\label{Taveragenonmon}
\end{eqnarray}
As we can see, the $1/D^2$ scaling inside the exponential persists, but because of the new term $C(\ell)=A \kappa(\ell)$, the mean escape time changes \emph{exponentially}. Two typical scenarios of nonmonotonic covariances are depicted in Figs. \ref{covnegative} and \ref{covpositive}. Shown in Fig. \ref{covnegative}a is the case when \emph{anti-correlations} of the disorder potential are present so that $\kappa(\ell)<0$. In this case the action (\ref{Snonmonotonic}) is \emph{smaller} than the action (\ref{actionresultgen}) obtained for $L\to \infty$, which leads to an exponential \emph{increase} of the particle escape time.
The physical mechanism of this increase is a much more frequent occurrence of large potential barriers in random realizations of the disorder. The optimal configuration of the disorder potential $V(x)$ in this case is shown in  Fig.  \ref{covnegative}b.

Figure \ref{covpositive}a depicts the case where  $\kappa(z)$ is nonmonotonic, but anti-correlations are absent, so that $\kappa(\ell)>0$. Here the action (\ref{Snonmonotonic}) is \emph{larger} than the action (\ref{actionresultgen}) obtained for $L\to \infty$, which leads to an exponential \emph{decrease} of the particle escape time.
Here large potential barriers in random realizations of the disorder are much more rare. The optimal configuration of the disorder potential $V(x)$ in this case is shown in  Fig.  \ref{covpositive}b.

\textit{Discussion.}  We found that regions of negative autocorrelations of one-dimensional disorder potential strongly (exponentially) suppress activated escape of overdamped particles, trapped in the local potential minima. In their turn, regions of nonmonotonic but positive autocorrelations strongly facilitate the escape.  A consequence of these results is a realization of the fact that quantitative modeling of one-dimensional particle transport in disordered media at low temperatures may require a more detailed knowledge of autocorrelation properties of the disorder than it was believed previously.

We assumed in our derivation that the covariance $\kappa(z)$ and the optimal configuration $V(x)$ are smooth functions of their arguments. It can be argued, however, that this assumption is actually unnecessary for Eq.~(\ref{probmonotone}) or (\ref{probnonmonotonic}) [and, as a result, Eq.~(\ref{Taveragemon}) or (\ref{Taveragenonmon}), respectively] to hold. To clarify this issue, we considered a well-known particular case of a non-smooth covariance: the Ornstein-Uhlenbeck potential, and did reproduce Eq.~(\ref{probmonotone}) leading to Eq.~(\ref{Taveragemon}). This calculation is presented in the Appendix.

A worthy next step is to address \emph{long-correlated} disorder, for which the integral $\int_{-\infty}^{\infty} C(z)\, dz$ diverges.

How can our prediction of the exponential increase (or decrease) of the
mean escape time due to negative (or positive but nonmonotonic) correlations of disorder, respectively, be
tested in experiment?
Colloidal systems in a speckled light field, where controllable disorder landscapes can be created \cite{G1,G3,G4,speckle},  look especially promising for this purpose. Unfortunately, particle mean-square displacement measurements alone are insufficient here.  The reason is that, remarkably, the effective long-time diffusion coefficient of particles is insensitive to autocorrelations of disorder and is determined solely by the disorder variance \cite{Zwanzig}.

I am indebted to Michael Wilkinson for introducing me to this problem and for useful discussions, and to Yael Roichman for a discussion of experiments with colloidal systems in a speckled light field.
I am especially grateful to Pavel Sasorov and Netanel Levi for bringing to my attention some errors in earlier versions of this work. Financial support for this research was provided in part by Grant No.
1499/20 from the Israel Science Foundation.

\appendix

\section*{Appendix. The Ornstein-Uhlenbeck disorder potential}

\renewcommand{\theequation}{A\arabic{equation}}
\setcounter{equation}{0}

The Ornstein-Uhlenbeck (OU) potential is a Gaussian random potential $V(x)$ with the covariance
\begin{equation}\label{covariance}
\langle V(x) V(x^\prime)\rangle = C(0) \,e^{-|x-x^\prime|/\ell},
\end{equation}
where $\ell$ is the correlation length.  This covariance has a corner singularity at $x=x^\prime$, so formally Eq.~(\ref{maximumcondition}) cannot be used. No less importantly, the barrier distribution that we are after is ill-defined in this case, as the potential $V(x)$ is not smooth, and its maxima and minima are everywhere dense. We argue, however, that \emph{very large} maxima and minima of the disorder potential are well behaved and amenable to the OFM. Therefore, we proceed as if we were unaware of the formal ill-posedness %\cite{regularize},
and solve the problem anew. As we will see shortly, the solution will again lead us to  Eqs.~(\ref{V2})-(\ref{Taveragemon}), as to be expected for a monotone-decreasing covariance.

The action functional of the OU process can be obtained from the general relation (\ref{correlatorgeneral}), and it is well known:
\begin{equation}\label{action1}
S[V(x)] = \frac{1}{4C(0)} \int_{-\infty}^{\infty} dx\,\left[\ell \left(\frac{dV}{dx}\right)^2+\frac{V^2}{\ell}\right],
\end{equation}
Let us minimize this functional subject to the condition (\ref{constraints}) on $\Delta V$.  The Euler-Lagrange equation reads
\begin{equation}\label{ELforOU}
\frac{d^2}{d x^2}V(x) - \frac{V(x)}{\ell^2} = 0.
\end{equation}
The spike-antispike solution of this equation is
\begin{equation}\label{V(x)OU}
V(x)=\pm \frac{\Delta V}{2} \,e^{-\frac{|x\mp L|}{\ell}}\,,
\end{equation}
where $L\to \infty$, in agreement with Eq. (\ref{V2}). Notice, that the maximum and minimum of $V(x)$ at $x=\pm L$ are not smooth, so Eqs.~(12) and (13) are inapplicable. But otherwise the optimal configuration of the disorder potential is similar to that for a smooth monotone-decreasing covariance. Furthermore, plugging the solution~(\ref{V(x)OU}) into Eq.~(\ref{action1}) and evaluating the integral, we reproduce Eqs.~(\ref{actionresultgen}) and (\ref{probmonotone}), leading to Eq.~(\ref{Taveragemon}).


\begin{thebibliography}{99}
\bibitem{otherconditions} As one can check, the second derivative of the optimal solution (\ref{V1}) at $x=L$
is negative as it should be, while the condition of the absence of additional extrema within the interval $|x|<L$ is obeyed automatically.
\bibitem{aboutLV} Lopatin and Vinokur \cite{LV} evaluated a slightly different distribution, which they call the delay time distribution, but it is simply related to $\mathcal{P}(\Delta V)$, and our result~(\ref{probmonotone}) for the monotonic covariance agrees with theirs.
\bibitem{limitation} We must demand that %$\kappa_k$,
the cosine Fourier transform of $\kappa(z)$ is nonnegative for all $k$. This is a necessary and sufficient condition for the action~(\ref{actiongeneral}) to be positive.
\end{thebibliography}
\end{document}